\documentclass[prd,onecolumn,nofootinbib,floatfix]{revtex4}
\usepackage{amssymb}
\usepackage{amsmath}
\usepackage{bm}

\newcommand{\xx}{\mathbf x}
\newcommand{\pp}{\mathbf p}
\newcommand{\kk}{\mathbf k}
\newcommand{\BB}{\mathbf B}
\newcommand{\bb}{\mathbf b}
\newcommand{\om}{\bm \omega}
\newcommand{\jj}{\mathbf j}

\begin{document}

\author{D.~Frenklakh}

\affiliation{B. Cheremushkinskaya, 25, Institute of Theoretical and Experimental Physics, Moscow 117218, Russia \\ Institutskii per, 9, Moscow Institute of Physics and Technology, Dolgoprudny 141700, Russia}

\title{Chiral heat wave and mixed waves in kinetic theory}

\begin{abstract}
We study collective excitations in hot rotating chiral media in presence of magnetic field in kinetic theory, namely Chiral Heat Wave and its' mixings with Chiral Vortical Wave and Chiral Magnetic Wave. Our results for velocities of these waves have slight alterations from those obtained earlier. We explain the origin of these alterations and also give the most general expressions for the velocities of all these waves in hydrodynamic approach.
\end{abstract}

\maketitle
\section{Introduction} \label{sec:Intro}

Transportation effects driven by anomalies in chiral systems have attracted large interest recently. They can be induced by external magnetic field like in the Chiral Magnetic Effect(CME) , where electric current is generated along the magnetic field due to the presence  of axial chemical potential(\cite{ref:CME}). In magnetic field there also exists the Chiral Separation Effect (CSE), where conversely axial current along the magnetic field is generated in presence of  vector chemical potential (\cite{ref:CSE1, ref:CSE2}). Another example is the Chiral Vortical Effect(CVE) (\cite{ref:CVE1,ref:CVE2,ref:CVE3,ref:CVE4}) occurring in rotating chiral systems. 

These anomalous transport effects couple vector and axial charge densities and currents which leads to the existence of gapless excitations such as the Chiral Magnetic Wave (CMW)(\cite{ref:CMW}) and the Chiral Vortical Wave (CVW)(\cite{ref:CVW}). The excitations existing in a hot rotating chiral system placed in external magnetic field were recently investigated in (\cite{ref:CHW}) in hydrodynamical framework. Chiral kinetic theory recently has been used to study transportation processes such as CME and CVE as well as the corresponding gapless excitations - CMW (\cite{ref:Kinetic1}), CVW (\cite{ref:CVW}) and mixed Chiral Magnetic-Vortical Wave (\cite{ref:Kin_my}) . In this paper we will use kinetic theory to rederive some of the results obtained in \cite{ref:CHW} and, as it turns out, to correct some of them in some way. To be more precise, kinetic theory shows that the expressions for the velocities of so-called mixed Heat-Vortical and Heat-Magnetic-Vortical waves as they are given in \cite{ref:CHW} are incomplete. Shortly, this is because in \cite{ref:CHW} were not taken into account the non-zero, in general, values of $\dfrac{\partial \epsilon}{\partial\mu_V}$ and $\dfrac{\partial \rho_V}{\partial T}$ (the notations here go as in \cite{ref:CHW}). In the framework of kinetic theory these values turn out to be proportional to the background vector chemical potential $\mu_V$, which is why the expressions for the velocities of Heat and mixed Heat-Magnetic waves derived through kinetic theory in this paper coincide with the earlier results in \cite{ref:CHW}. However, there is a room for discussion on generality of this proportionality to $\mu_V$ and, until proven useless, it might be right to write these expressions in the most general way. 

\section{General discussion in kinetic theory}\label{sec:General}

We study a system of right and left Weyl fermions, denoted by indices R and L respectively. There are particles and antiparticles of each kind, and we denote them by indices + and $-$, so that R+ are right particles, R$-$ are antiparticles to them and analogously for left (this notation may be a bit confusing because antiparticles to right particles are actually left). We will always consider a rotating system, with angular velocity denoted by $\om$. Sometimes we will also need external magnetic field $\BB$ and non-zero background vector chemical potential $\mu$. The axial chemical potential is always set to zero in the background. The temperature of the system is denoted by $T$ and is large compared to all the other parameters which are $\omega$, $\mu$ and $\sqrt{B}$ (see \cite{ref:Kinetic1}).

Taking similar approach as  in(\cite{ref:CVW, ref:Kinetic1})
we start from the kinetic equation
\begin{equation}\label{kinetic}
\frac{\partial f_{\pm R/L}}{\partial t} + \dot\xx\cdot\frac{\partial f_{\pm R/L}}{\partial \xx}+\dot\pp\cdot\frac{\partial f_{\pm R/L}}{\partial \pp} = C_{\pm R/L} [f_{+R/L},f_{-R/L}] ,
\end{equation}
where $f_{\pm R/L}(t,\xx,\pp)$ are distribution functions and $C_{\pm R/L}[f_+,f_-]$ are collision integrals. 
For the general discussion let us consider the most general case with magnetic field turned on. Then the equations of motion for right particles and their antiparticles $(R\pm)$ in their local rest frame are
\begin{equation}
\dot\xx = \hat\pp + \dot\pp\times\bb , ~\dot\pp = \dot\xx\times\BB_{\pm}' ,
\end{equation}
Here $p = |\pp|$, $\hat\pp = \dfrac{\pp}{p}$ and $\bb = \dfrac{\hat\pp}{2 p^2}$ is the curvature of Berry's connection in momentum space (see \cite{ref:Kinetic2} for details); $\BB_{\pm}' = \BB \pm 2p~\om$ denote the effective magnetic field. The equations for $L\pm$ are the same, just $\bb$ should be replaced with $-\bb$. From these equations we obtain

\begin{eqnarray}
\sqrt{G_{\pm R}}\dot\xx = \hat\pp + \frac{\BB_{\pm}'}{2p^2}  ,\\
\sqrt{G_{\pm L}}\dot\xx = \hat\pp - \frac{\BB_{\pm}'}{2p^2}  ,\\
\sqrt{G_{\pm R}}\dot\pp = \sqrt{G_{\pm L}}\dot\pp = \pm \pp\times\BB_{\pm}' , 
\end{eqnarray}
where $\sqrt{G_{\pm R}} = 1 + \dfrac{\BB'_{\pm}\cdot\om}{p^2}$ and $\sqrt{G_{\pm L}} = 1 - \dfrac{\BB'_{\pm}\cdot\om}{p^2}$ modify phase space volume due to the interplay between effective magnetic field and Berry connection.

We now want to linearise the kinetic equation. Consider small fluctuations above the equilibrium Fermi-Dirac distribution $f_{0\pm R/L}(p)$:
\begin{equation}
f_{\pm R/L}=f_{0\pm R/L}(p)-\partial_p f_{o\pm R/L}(p)\delta f_{\pm R/L}(t,\xx,\pp) ,
\end{equation}
and take the Fourier transformation of $\delta f_{\pm R/L}(t,\xx,\pp)$ to be $h_{\pm R/L}(\nu,\kk,\pp)$. In equilibrium the collision term is zero, so we can write it as
\begin{equation}
C_{\pm R/L}[f_{+R/L},f_{-R/L}] = -\partial_p f_{0\pm R/L}I_{\pm R/L}[h_{+R/L},h_{-R/L}]+O(h^2) .
\end{equation}
The kinetic equation (\ref{kinetic}) now becomes 
\begin{equation}\label{reduced}
-i\nu h_{\pm R/L} + \dot\xx(i\kk \pm \BB_{\pm}\times\frac{\partial}{\partial\pp})h_{\pm R/L} = I_{\pm R/L}[h_{+R/L},h_{-R/L}] .
\end{equation}

Now we want to get rid of collision integrals for which we exploit the conservation of some quantities. We average (\ref{reduced}) in the momentum space with the brackets $\langle...\rangle_{\pm R/L}$ defined as
\begin{equation}
\langle...\rangle_{\pm R/L} = \int_{\pp}\sqrt{G_{\pm R/L}}\partial_p f_{0\pm R/L}(p)(...) ,
\end{equation}
 where $\int_{\pp} = \int\dfrac{d^3p}{(2\pi)^3}$. From the conservation of the total number of right or left particles ($\{R/L+\} - \{R/L-\}$) one obtains  
\begin{equation}
\int_{\pp}\sqrt{G_{+ R/L}}C_{+ R/L}[f_+,f_-]-\int_{\pp}\sqrt{G_{-R/L}}C_{-R/L}[f_+,f_-] = 0 ,
\end{equation}
for arbitrary $f_{\pm R/L}$ which implies 
\begin{equation}
\int_{\pp}\sqrt{G_{+R/L}}\partial_p f_{0+ R/L} I_{+R/L}[h_{\pm R/L}] - \int_{\pp}\sqrt{G_{-R/L}}\partial_p f_{0- R/L} I_{-R/L}[h_{\pm R/L}] = 0 ,
\end{equation}
for arbitrary $h_{\pm R/L}$.
Also, the "Lorentz force" term vanishes after averaging and integrating by parts.
So, averaging the equations (\ref{reduced}) for right/left particles and antiparticles with the corresponding bracket and taking the difference we obtain
\begin{eqnarray}\label{empty}
\nu(\langle h_{+R}\rangle_{+R} - \langle h_{-R}\rangle_{-R}) - \kk(\langle\dot\xx h_{+R}\rangle_{+R} - \langle\dot\xx h_{-R}\rangle_{-R}) = 0 , \\
\nu(\langle h_{+L}\rangle_{+L} - \langle h_{-L}\rangle_{-L}) - \kk(\langle\dot\xx h_{+L}\rangle_{+L} - \langle\dot\xx h_{-L}\rangle_{-L}) = 0 .
\end{eqnarray}

Another conservation to exploit is energy conservation. For using it we  multiply (\ref{reduced}) by $p$, average with the corresponding bracket and take the sum over $\pm$ and R/L. The collision integral terms will vanish due to energy conservation so one obtains

\begin{equation} \label{en_empty}
\nu(\langle p~ h_{+R}\rangle_{+R} + \langle p~ h_{-R}\rangle_{-R} + \langle p~ h_{+L}\rangle_{+L} + \langle p~ h_{-L}\rangle_{-L}) - \kk(\langle p~ \dot\xx~ h_{+R}\rangle_{+R} + \langle p ~\dot\xx ~h_{-R}\rangle_{-R} +\langle p ~\dot\xx~ h_{+L}\rangle_{+L} + \langle p~ \dot\xx~ h_{-L}\rangle_{-L}) = 0 .
\end{equation}

In the next few sections we will study equations (\ref{empty}-\ref{en_empty}) in hydrodynamic regime under different conditions. There are two possible ways : one is to derive the most general expression first and then explore different special cases, while the other is to start with the easiest case possible and complicate things gradually. We will use the second way, as it will allow us to see clearly, where the difference between the results obtained in kinetic theory and in paper \cite{ref:CHW} starts to emerge. 

\section{Different waves in hydrodynamic regime in kinetic theory}\label{sec:Waves}

\subsection{Pure heat wave}\label{subsec:H}

Now we consider the case $\mu_V = \mu_A = \mu_L = \mu_R = 0$ in background and magnetic field is turned off : $\BB = 0$, in this case only heat wave exists. We want to study hydrodynamic regime in which the fluctuations above the equilibrium correspond to infinitesimal shifts in chemical potential and temperature. We will consider the shifts of chemical potentials to be $\delta \mu_R = -\delta\mu_L \equiv \delta\mu$ - as we will see in the next subsection, this would be just enough for the case of the pure heat wave. Equilibrium distribution is Fermi-Dirac distribution $f_{0\pm R/L} = \dfrac{1}{e^{\beta p} + 1}$. The corresponding shift of the equilibrium distribution function is $\delta f_{+R} = -\partial_p f_0(p)\beta\delta\mu = -\delta f_{-R} = -\delta f_{L+} = \delta f_{L-}$. For the shift of temperature $\delta\beta$ the corresponding shift of distribution function is $\delta f_{\pm R/L} = \partial_\beta f_0 \delta\beta = \partial_p f_0 \dfrac{p}{\beta} \delta\beta$. To make the notations look as in the previous section we now change $-\beta\delta\mu\to h_1$ and $\dfrac{\delta\beta}{\beta^2}\to h_2$. The resulting fluctuation in distribution function is now parametrised as 
\begin{eqnarray}
\delta f_{R+} = \partial_p f_0(p) (h_1 + \beta p ~h_2) = \delta f_{L-} ,\\
\delta f_{R-} = \partial_p f_0(p) (-h_1 + \beta p~ h_2) = \delta f_{L+}.
\end{eqnarray}
Now we plug this into (\ref{empty}) and (\ref{en_empty})
\begin{eqnarray}
\nu(\langle h_1+\beta p ~h_2\rangle_{+R} - \langle -h_1+\beta p ~h_2\rangle_{-R}) + \kk(\langle \dot\xx(h_1+\beta p ~h_2)\rangle_{+R} - \langle \dot\xx(-h_1+\beta p ~h_2)\rangle_{-R})=0 ,\\
\nu(\langle p~h_1 + \beta p^2~h_2\rangle_{+R} + \langle -p~h_1 +\beta p^2 h_2\rangle_{-R}) + \kk(\langle\dot\xx(p~h_1 + \beta p^2~h_2)\rangle_{+R}) + \langle\dot\xx(-p~h_1 + \beta p^2~h_2)\rangle_{-R}) = 0.
\end{eqnarray}
Let us denote $\sqrt{G_{R+}} = \sqrt{G_{L-}} \equiv G_1$ and $\sqrt{G_{R-}} = \sqrt{G_{L+}}\equiv G_2$ for future convenience. Rewriting the  above equations:
\begin{eqnarray}
\nu\int_\pp\partial_p f_0[(G_1+G_2)h_1 + \beta p (G_1 - G_2)h_2] + \kk\int_\pp \partial_p f_0[(G_1\dot\xx + G_2\dot\xx)h_1 + \beta p(G_1\dot\xx - G_2\dot\xx)h_2] = 0 ,\\
\nu\int_\pp\partial_p f_0[p(G_1-G_2)h_1 + \beta p^2(G_1+G_2)h_2 ] + \kk\int_\pp\partial_p f_0[p(G_1\dot\xx - G_2\dot\xx)h_1 + \beta p^2(G_1\dot\xx + G_2\dot\xx)h_2] = 0. 
\end{eqnarray}

Note that $\int_\pp \pp g(p) = 0$ for arbitrary $g(p)$, so for our integrals $G_{1,2}\sim1$ and $G_{1,2}\dot\xx\sim\pm\dfrac{\om}{p}$. Then the last two equations are easily rewritten as 

\begin{equation}\label{simple_system_1}
2\nu h_1\int_\pp\partial_p f_0 + 2\beta(\om\cdot\kk) h_2\int_\pp\partial_p f_0 = 0,
\end{equation}
\begin{equation}\label{simple_system_2}
2\nu\beta h_2\int_\pp p^2\partial_p f_0 + 2(\om\cdot\kk) h_1\int_\pp\partial_p f_0 = 0.
\end{equation}

In future it would be convenient to see the physical meaning of the integrals in the last equations:
\begin{equation}
 2\int_\pp\partial_p f_0 = -\int_\pp(G_1+G_2)\frac{\partial f_0}{\partial\mu} = -\frac{\partial n_R}{\partial\mu}\equiv-\chi, 
\end{equation} 
 where $n_R$ is charge density of right particles and $\chi$ is charge susceptibility. On the other hand,
\begin{equation}
\int_\pp\partial_p f_0 = \frac{1}{2\pi^2}\int_0^\infty p^2\partial_p f_0 dp = -\frac{1}{\pi^2}\int_0^\infty \frac{p~ dp}{e^{\beta p}+1} = -\frac{T^2}{12}
\end{equation} 
Also, 
\begin{equation}
\int_\pp p^2\partial_p f_0 = \beta\int_\pp p~\partial_\beta(\frac{1}{e^{\beta p}+1}) = \beta\int_\pp\partial_\beta(f_0 p) = \frac{1}{4}\beta\partial_\beta\int_pp (2G_1 + 2G_2)p~f_0 = \frac{\beta}{4}\partial_\beta\epsilon  = \frac{-TC_V}{4},
\end{equation} 
where $\epsilon$ is energy density and $C_V$ is heat capacity. And again on the other hand:
\begin{equation}
\int_\pp p^2 \partial_p f_0 = - \frac{2}{\pi^2}\int_0^\infty\frac{p^3 dp}{e^{\beta p} + 1} = -\frac{7\pi^2}{30}T^4 ,
\end{equation}
Using these relations, we rewrite equations (\ref{simple_system_1}, \ref{simple_system_2}) as
\begin{eqnarray}
\nu h_1 + \beta(\om\cdot\kk)h_2=0 , \\
\nu \frac{C_V}{2} h_2 + (\om\cdot\kk)\chi h_1 = 0.
\end{eqnarray}

Excluding one of $h'$s from this system we easily obtain the dispersion relation for $\kk||\om$:
\begin{equation}
\nu = v_{CHW}k,
\end{equation}
where 
\begin{equation}
v_{CHW} = \omega\sqrt{\frac{2\chi}{C_VT}}.
\end{equation}
Using the relation $\chi = \dfrac{T^2}{6}$, we can transform it to the form
\begin{equation} \label{HeatWave}
v_{CHW} = \frac{\omega}{3}\sqrt{\frac{T^3}{2C_V\chi}},
\end{equation}
which coincides with \cite{ref:CHW} if one takes into account that in that paper $\chi$ differs by the factor of $2$ from the definition in this paper.

It might also be interesting to express the velocity in terms of only temperature and angular velocity, as they are the only independent parameters here:
\begin{equation}
v_{CHW} = \frac{\omega}{\pi T}\sqrt{\frac{5}{14}}
\end{equation}

\subsection{Mixed heat and vortical wave}\label{subsec:HV}
Now we consider the case $\mu_V = \mu_R=\mu_L\equiv\mu\neq0$, $\BB=0$. In this case background equilibrium distributions are $f_{0R+} = f_{0L+} = \dfrac{1}{e^{\beta(p-\mu)} + 1}\equiv f_{0+}$ and $f_{0R-} =  f_{0L-} = \dfrac{1}{e^{\beta(p+\mu)}+1}\equiv f_{0-}$. We want to study hydrodynamic regime again, but now we don't impose any constraints on the fluctuation of $\mu_V$, as before. So, now we have 3 independent fluctuations : $\delta\mu_R$, $\delta\mu_L$ and $\delta\beta$. The corresponding fluctuations in distribution functions are 
\begin{eqnarray}
\delta f_{+ R/L} = \partial_p f_{0+} (-\beta\delta\mu_{R/L}+ \frac{p-\mu}{\beta}\delta\beta) ,\\
\delta f_{-R/L} = \partial_p f_{0-} (\beta\delta\mu_{R/L} + \frac{p+\mu}{\beta}\delta\beta) .
\end{eqnarray}
Denoting, like before, $-\beta\delta\mu_R - \dfrac{\mu}{\beta}\delta\beta$ as $h_1$, $-\beta\delta\mu_L + \dfrac{\mu}{\beta}\delta\beta$ as $h_2$ and $\dfrac{\delta\beta}{\beta^2}$ as $h_3$ and plugging it back to (\ref{empty}) - (\ref{en_empty}), we get
\begin{eqnarray}
\nonumber
\nu\int_\pp[\partial_p f_{0+}G_1(h_1+\beta p~h_3) - \partial_p f_{0-}G_2(-h_1+\beta p~h_3)] + \kk\int_\pp[\partial_p f_{0+}G_1\dot\xx(h_1+\beta p~h_3) - \partial_p f_{0-}G_2\dot\xx(-h_1+\beta p~h_3)], \\
\nu\int_\pp[\partial_p f_{0+}G_2(h_2+\beta p~h_3) - \partial_p f_{0-}G_1(-h_2+\beta p~h_3)] + \kk\int_\pp[\partial_p f_{0+}G_2\dot\xx(h_2+\beta p~h_3) - \partial_p f_{0-}G_1\dot\xx(-h_2+\beta p~h_3)], \\ \nonumber
\nu\int_\pp(\partial_p f_{0+}G_1p(h_1+\beta p~h_3) + \partial_p f_{0-}G_2p(-h_1+\beta p~h_3)+\partial_p f_{0+}G_2p(h_2+\beta p~h_3) + \partial_p f_{0-}G_1p(-h_2 +\beta p~h_3)) + \\ \nonumber
 \kk\int_\pp(\partial_p f_{0+}G_1\dot\xx p(h_1+\beta p~h_3) + \partial_p f_{0-}G_2\dot\xx p(-h_1+\beta p~h_3) + \partial_p f_{0+}G_2\dot\xx p(h_2 + \beta p~h_3) + \partial_p f_{0-}G_1\dot\xx p(-h_2 + \beta p~h_3)).
\end{eqnarray}
From these equations using definitions of $G_{1,2}$ and the equations of motion as well as the integrals
\begin{eqnarray}
\int_\pp(\partial_p f_{0+} + \partial_p f_{0-}) = -\chi, \\
\int_\pp 2p^2(\partial_p f_{0+}+\partial_p f_{0-}) = -TC_V,\\
\int_\pp p (\partial_p f_{0+} - \partial_p f_{0-}) = -\frac{\mu T^2}{2}, \\
\int_\pp \frac{1}{p} (\partial_p f_{0+} - \partial_p f_{0-}) = -\frac{\mu}{2\pi^2},
\end{eqnarray}
we may transform the system to the form
\begin{equation}
\left[\nu\begin{pmatrix}
\chi & 0 &\mu T/2 \\ 0 & \chi & \mu T/2 \\ \mu T/2 & \mu T/2 & C_V/T
\end{pmatrix}
+(\kk\cdot\om)\begin{pmatrix}
\mu/2\pi^2 & 0 & \chi/T \\ 0 & -\mu/2\pi^2 & -\chi/T \\ \chi/T & -\chi/T & 0
\end{pmatrix}
\right] \begin{pmatrix}
h_1 \\ h_2 \\ h_3
\end{pmatrix} = \begin{pmatrix}
~0~ \\ ~0~ \\ ~0~
\end{pmatrix}.
\end{equation}

At this point it is clear, that if $\mu=0$, than $h_1=-h_2$, as we supposed in the previous subsection. By introducing new variables $h_V = h_1+ h_2 $, $h_A = h_1-h_2$ and leaving $h_3$ as it was, we come to the new system

\begin{equation}
\left[\nu\begin{pmatrix}
\chi & 0 & \mu T \\ 0 & \chi & 0 \\ \mu T/2 & 0 & C_V/T
\end{pmatrix}
+ (\kk\cdot\om)\begin{pmatrix}
0 & \mu/2\pi^2 & 0 \\ \mu/2\pi^2 & 0 & 2\chi/T \\ 0 & \chi/T & 0
\end{pmatrix}
\right] \begin{pmatrix}
h_V \\ h_A \\ h_3
\end{pmatrix} = \begin{pmatrix}
~0~ \\ ~0~ \\ ~0~
\end{pmatrix} .
\end{equation}

Multiplying this equation by the inverse of the first matrix and keeping the terms only up to the second order in $\mu$ as we work in the approximation of small $\mu$ we get

\begin{equation} \label{mat_fin_hv}
\left[\nu\begin{pmatrix}
1 & 0 & 0 \\ 0  & 1 & 0 \\ 0 & 0 & 1
\end{pmatrix} + (\kk\cdot\om)\begin{pmatrix}
0 & \mu/(2\pi^2\chi) - \mu T/(2C_V) & 0 \\ \mu/(2\pi^2\chi) & 0 & 2/T \\ 0 & \chi/C_V- \mu^2 T^2/(2\pi^2C_V\chi) + \mu^2T^3/2C_V^2 & 0
\end{pmatrix}
\right] \begin{pmatrix}
h_V \\ h_A \\ h_3
\end{pmatrix} = \begin{pmatrix}
~0~ \\ ~0~ \\ ~0~
\end{pmatrix} .
\end{equation}
So, the coefficient in linear dispersion relation is determined by the eigenvalues of the second matrix in  (\ref{mat_fin_hv}) which are $0$ and $\pm v_{CHVW}$, where

\begin{equation}
v_{HV} = \omega\sqrt{\frac{2\chi}{TC_V} - \frac{\mu^2T}{2\pi^2C_V\chi} + \frac{\mu^2T^2}{C_V^2} + \frac{\mu^2}{4\pi^4\chi^2} }.
\end{equation}

We would like to stress that this result is valid up to terms quadratic in $\mu$, as we neglected the terms of higher order while obtaining it. To make it look closer to the \cite{ref:CHW} result we note, that, with a quadratic in $\mu$ term taken into account, $\chi\approx \dfrac{T^2}{6} + \dfrac{\mu^2}{4\pi^2}$,  so 

\begin{equation}
v_{HV} = \omega\sqrt{\frac{T^3}{18 C_V\chi} + \mu^2(\frac{1}{4\pi^4\chi^2} + \frac{T}{6\pi^2C_V\chi} + \frac{T^2}{C_V^2} - \frac{5T}{4\pi^2 \chi C_V})}.
\end{equation}

The last two terms are not present in \cite{ref:CHW}. This is because there weren't taken into account the off-diagonal terms in the matrix accompanying $\nu$. Indeed, it is easy to check, that with these terms set to zero, the result would be the same as in \cite{ref:CHW}. These terms correspond to the non-zero quantities $\dfrac{\partial \epsilon}{\partial\mu_V}$ and $\dfrac{\partial \rho_V}{\partial T}$, where $\epsilon$ is energy density and $\rho_V$ is vector charge density. As it follows from the above calculations, these terms are proportional to $\mu_V$ and that is why they didn't arise in the case of pure heat wave.$•$

\subsection{Mixed heat and magnetic wave}\label{subsec:HM}

Now let us consider the case of $\mu = 0$, $\BB\neq0$. From the discussion above we expect the off-diagonal  terms in the matrix accompanying $\nu$ to vanish and therefore we might predict, that our result in this section would be the same as in \cite{ref:CHW}. As we will see, this is a correct prediction.

 As before in the case of pure heat wave, we have equilibrium distribution function $f_0 = \dfrac{1}{e^{\beta p} +1}$ and we are looking for its' fluctuations parametrised as $\delta f_{R\pm} = \dfrac{\partial f_0}{\partial p}(\pm h_1 +\beta p~h_3)$ and $\delta f_{L\pm} = \dfrac{\partial f_0}{\partial p}(\pm h_2 + \beta p~h_3)$. The modified phase space volumes are now $\sqrt{G_{R+}} = 1+ \BB_+'\cdot\bb\equiv G_1$, $\sqrt{G_{R-}} = 1 + \BB_-'\cdot\bb\equiv G_2$, $\sqrt{G_{L+}} = 1 - \BB_+'\cdot\bb\equiv G_3$, $\sqrt{G_{L-}} = 1- \BB_-'\cdot\bb\equiv G_4$. So, plugging it into (\ref{empty}) - (\ref{en_empty}), we get
 
 \begin{eqnarray}
\nonumber
\nu\int_\pp\partial_p f_0[G_1(h_1+\beta p~h_3) - G_2(-h_1+\beta p~h_3)]+\kk\int_\pp\partial_p f_0 [G_1\dot\xx(h_1+\beta p~h_3) - G_2\dot\xx(-h_1+\beta p~h_3)] = 0 ,\\
\nonumber
\nu\int_\pp\partial_p f_0[G_3(h_2+\beta p~h_3) - G_4(-h_2+\beta p~h_3)]+\kk\int_\pp\partial_p f_0 [G_3\dot\xx(h_2+\beta p~h_3) - G_4\dot\xx(-h_2+\beta p~h_3)] = 0 ,\\
\nu\int_\pp\partial_p f_0 p~[G_1(h_1+\beta p~h_3) + G_2(-h_1+\beta p~h_3)+G_3(h_2+\beta p~h_3) + G_4(-h_2+\beta p~h_3)] + \\
\nonumber
\kk\int_\pp\partial_p f_0 p~[G_1\dot\xx(h_1+\beta p~h_3) + G_2\dot\xx(-h_1+\beta p~h_3) + G_3\dot\xx(h_2+\beta p~h_3) + G_4\dot\xx(-h_2+\beta p~h_3)] = 0.
 \end{eqnarray}
From these equations after using the equations of motion along with the integrals that were already used in section (\ref{subsec:H}) we obtain the system
\begin{equation}
\left[\nu\begin{pmatrix}
\chi & 0 & 0 \\ 0 & \chi & 0 \\ 0 & 0 & C_V/T
\end{pmatrix} + \begin{pmatrix}
(\kk\cdot\BB)/(4\pi^2) & 0 & T/6(\kk\cdot\om) \\ 0 & (\kk\cdot\BB)/(4\pi^2)& -T/6(\kk\cdot\om) \\ T^2/6(\kk\cdot\om) & -T^2/6(\kk\cdot\om) & 0
\end{pmatrix}\right]\begin{pmatrix}
h_1 \\ h_2 \\ h_3
\end{pmatrix} = \begin{pmatrix}
0 \\ 0 \\ 0
\end{pmatrix}.
\end{equation}

Introducing, as before, new variables $h_V = h_1+ h_2 $, $h_A = h_1-h_2$ we get
\begin{equation}
\left[\nu\begin{pmatrix}
1 & 0 & 0 \\ 0 & 1 & 0 \\ 0 & 0 & 1
\end{pmatrix} + \begin{pmatrix}
0 & (\kk\cdot\BB)/(4\pi^2\chi) & 0 \\ (\kk\cdot\BB)/(4\pi^2\chi) & 0 & T^2/(6\chi)(\kk\cdot\om) \\ 0 & T^2/(3C_V)(\kk\cdot\om) & 0
\end{pmatrix}\right] \begin{pmatrix}
h_V \\ h_A \\ h_3
\end{pmatrix} = \begin{pmatrix}
0 \\ 0 \\ 0
\end{pmatrix}.
\end{equation}

The values of $\nu$ corresponding to a given $\kk$ are simply eigenvalues of the second matrix, which are $0$ and $\pm\sqrt{\dfrac{(\kk\cdot\BB)^2}{16\pi^4\chi^2} + \dfrac{T^3}{18\chi C_V}(\kk\cdot\om)^2}$. We want to find the maximal and minimal possible relation $\nu/k$ for $\kk$ lying in the same plane as $\BB$ and $\om$ (then, similar to \cite{ref:CHW} we will be able to find the speed of the wave for arbitrary $\kk$ ). Let us introduce the following angles: $\phi$ between $\BB$ and $\om$, $\phi_1$ between $\kk$ and $\om$ and $\phi_2$ between $\kk$ and $\BB$ (so $\phi_2 = \phi - \phi_1$). Then for non-trivial $\nu$ we have
\begin{equation}
(\frac{\nu}{k})^2 = \frac{B^2}{16\pi^4\chi^2}\cos^2\phi_1 + \frac{T^3\omega^2}{18\chi C_V}\cos^2\phi_2.
\end{equation}
The maximum and minimum of this expression lie at the points where $\tan2\phi_1 = \dfrac{v_{CHW}^2\sin2\phi}{v_{CMW}^2 + v_{CHW}^2\cos2\phi}$ and in these points they are equal
\begin{equation}
v_{MH\pm}^2 = \frac{v_{CHW}^2+v_{CMW^2}}{2} \pm \frac{1}{2}\sqrt{v_{CHW}^4 + v_{CMW}^4 + 2v_{CHW}^2v_{CMW}^2\cos2\phi}.
\end{equation}
Here we used the expression for the velocity of pure heat wave (\ref{HeatWave}) and the expression for the velocity of Chiral Magnetic Wave $v_{CMW} = \dfrac{B}{4\pi^2\chi}$ (\cite{ref:CMW}, \cite{ref:Kinetic1}). As we predicted, this answer is the same as in \cite{ref:CHW}, so we reobtained this result through kinetic theory. Let us notice, that the larger velocity is never equal to zero, so there always exists a propagating mode.

\subsection{Mixed heat, magnetic and vortical wave}\label{subsec:HMV}

Now we finally consider the case $\BB\neq0$, $\mu\neq0$ in which all the three waves should mix. The background distribution functions are again $f_{0R+} = f_{0L+} = f_{0+} = \dfrac{1}{e^{\beta(p-\mu)}+1}$, $f_{0R-} = f_{0L-} = f_{0-} = \dfrac{1}{e^{\beta(p+\mu)}+1}$ and we parametrise the fluctuations above them again as $\delta f_{R\pm} = \dfrac{\partial f_{0\pm}}{\partial p}(\pm h_1 +\beta p~h_3)$ and $\delta f_{L\pm} = \dfrac{\partial f_{0\pm}}{\partial p}(\pm h_2 + \beta p~h_3)$ (like in \ref{subsec:HV}).
Using the notations $G_i ( i = \overline{1,4})$ from subsection \ref{subsec:HM} and plugging everything in (\ref{empty}-\ref{en_empty}), we get

\begin{eqnarray}
\nonumber
\nu\int_\pp[\partial_p f_{0+}G_1(h_1+\beta p~h_3) - \partial_p f_{0-}G_2(-h_1+\beta p~h_3)]+\kk\int_\pp[\partial_p f_{0+} G_1\dot\xx(h_1+\beta p~h_3) - \partial_p f_{0-}G_2\dot\xx(-h_1+\beta p~h_3)] = 0 ,\\
\nonumber
\nu\int_\pp[\partial_p f_{0+}G_3(h_2+\beta p~h_3) - \partial_p f_{0-} G_4(-h_2+\beta p~h_3)]+\kk\int_\pp[\partial_p f_{0+}G_3\dot\xx(h_2+\beta p~h_3) -\partial_p f_{0-} G_4\dot\xx(-h_2+\beta p~h_3)] = 0 ,\\
\nu\int_\pp p(\partial_p f_{0+}[G_1(h_1+\beta p~h_3) + G_3(h_2+\beta p~h_3)] + \partial_p f_{0-} [G_2(-h_1+\beta p~h_3) + G_4(-h_2+\beta p~h_3)]) + ~~~~~~\\
\nonumber
\kk\int_\pp p(\partial_p f_{0+}[G_1\dot\xx(h_1+\beta p~h_3) + G_3\dot\xx(h_2+\beta p~h_3) + \partial_p f_{0-} G_2\dot\xx(-h_1+\beta p~h_3) + G_4\dot\xx(-h_2+\beta p~h_3)] = 0.
\end{eqnarray}
 Using he equations of motion and integrals that were calculated earlier and introducing variables $h_V = h_1 + h_2$ and $h_A = h_1-h_2$ we obtain
 \begin{equation} \label{MatrixHMV}
 \left[\nu\begin{pmatrix}
 \chi & 0 &\mu T \\ 0 & \chi & 0 \\ \mu T/2 & 0 & C_V/T
 \end{pmatrix} + \begin{pmatrix}
 0 & \dfrac{(\kk\cdot\BB) + 2\mu(\kk\cdot\om)}{4\pi^2} & 0 \\ \dfrac{(\kk\cdot\BB) + 2\mu(\kk\cdot\om)}{4\pi^2} & 0 & \dfrac{\mu(\kk\cdot\BB) + 4\pi^2\chi(\kk\cdot\om)}{2\pi^2T}\\ 0& \dfrac{\mu(\kk\cdot\BB) + 4\pi^2\chi(\kk\cdot\om)}{4\pi^2T} & 0
 \end{pmatrix}\right] \begin{pmatrix}
 h_V \\ h_A \\ h_3
 \end{pmatrix} = \begin{pmatrix}
 0 \\ 0 \\ 0
 \end{pmatrix}
 \end{equation}
As in (\ref{subsec:HV}) we multiply the equation by the inverse of the matrix accompanying $\nu$ and keep only terms up to the quadratic order in $\mu$. We obtain that there are three branches of dispersion relation looking like $\nu = 0, \pm\lambda$ where $\lambda$ is the non-zero positive eigenvalue (if they are not all zero) of the matrix
\begin{equation}
\begin{pmatrix}
0 & \dfrac{(\kk\cdot\BB')}{4\pi^2\chi}+\dfrac{\mu^2(T^3-\chi T)(\kk\cdot\BB)}{8\pi^2\chi^2C_V}-\dfrac{\mu T (\kk\cdot\om)}{2C_V} & 0 \\ \dfrac{(\kk\cdot\BB')}{4pi^2\chi} & 0 & \dfrac{\mu(\kk\cdot\BB)}{2\pi^2 T\chi} + \dfrac{2(\kk\cdot\om)}{T} \\ 0 & -\dfrac{\mu T^2(\kk\cdot\BB')}{4\pi^2\chi C_V} +\dfrac{\mu(\kk\cdot\BB)}{4\pi^2C_V}+\dfrac{(2\chi C_V+\mu^2T^3)(\kk\cdot\om)}{2C_V^2} & 0
\end{pmatrix},
\end{equation}
which is
\begin{equation}
\sqrt{\frac{(\kk\cdot\BB')}{4\pi^2\chi}\left[\frac{(\kk\cdot\BB')}{4\pi^2\chi} + \frac{5\mu^2T^3(\kk\cdot\BB)}{48\pi^2\chi^2C_V} - \dfrac{\mu T(\kk\cdot\om)}{2C_V} \right] + \left[\frac{\mu(\kk\cdot\BB)}{2\pi^2T\chi}+\frac{2(\kk\cdot\om)}{T}\right]\left[\frac{\mu(\kk\cdot\BB)}{4\pi^2C_V} - \frac{\mu T^2(\kk\cdot\BB')}{4\pi^2\chi C_V}+\frac{(2C_V\chi + \mu^2T^3)(\kk\cdot\om)}{2C_V^2}\right]}
\end{equation}
Here $\BB' = \BB + 2\mu\om$. Again, we are interested in the maximal and minimal possible relation $\nu/k$ for $\kk$, $\BB$ and $\om$ all lying in the same plane and we introduce angles as in \ref{subsec:HM}. Then the expression for $(\lambda/k)^2$ has the form

\begin{equation}\label{cosHMV}
\frac{\lambda^2}{k^2} = a_1 \cos^2\phi_1 + a_2 \cos\phi_1\cos\phi_2 + a_3\cos^2\phi_2
\end{equation}
with the coefficients
\begin{eqnarray}\label{eq:coefHMV}
\nonumber
a_1 = \frac{B^2}{16\pi^4\chi^2}\left(1+\frac{5\mu^2T}{6\chi C_V}\right), ~~~~~~~~~~~~~~~~~~~\\
a_2 = \frac{B\mu\omega}{2\pi^2}\left(\frac{1}{2\pi^2\chi^2} - \frac{T}{4C_V\chi} + \frac{2}{TC_V} + \frac{\mu^2T}{8\pi^2\chi^2C_V} + \frac{\mu^2T^2}{2\chi C_V^2}\right), \\ \nonumber
a_3 = \omega^2\left(\frac{2\chi}{TC_V} + \mu^2\left[\frac{1}{4\pi^2\chi^2} - \frac{5T}{\pi^2\chi C_V} + \frac{T^2}{C_V^2}\right]\right) = v_{CHVW}^2.
\end{eqnarray}
The maximum and minimum of the expression in (\ref{cosHMV}) are
\begin{equation} \label{vHMV}
v_{HMV\pm}^2 = \frac{a_1 + a_2\cos\phi+a_3}{2} \pm \frac{1}{2}\sqrt{a_1^2 + a_2^2 + a_3^2 +2a_2(a_1+a_3)\cos\phi+2a_1a_3\cos2\phi},
\end{equation}
and they are reached at the points where
\begin{equation}
\tan2\phi_1 = \frac{a_2\sin\phi + a_3\sin2\phi}{a_1+a_2\cos\phi + a_3\cos2\phi}.
\end{equation}
Now it is not so easy to compare this to the result in \cite{ref:CHW}, but we know                  what to expect : if there were no off-diagonal terms in the matrix accompanying $\nu$ in (\ref{MatrixHMV}) then the answer would be the same as in \cite{ref:CHW}. It is easy to check, that it is true. So, once again, the result obtained in \cite{ref:CHW} should be corrected due to the non-zero values of $\dfrac{\partial\rho_V}{\partial T}$ and $\dfrac{\partial \epsilon}{\partial \mu_V}$. Let us notice that the expression for $v_{CHMVW\pm}$ with the coefficients (\ref{eq:coefHMV}) is always larger than zero, as we work in the case $\mu\ll T$ so the larger velocity is never zero and there is always a propagating mode. G

\section{Hydrodynamic approach}\label{sec:hydro}

In this section we take the same approach as in \cite{ref:CHW} but we take into account the non-zero values of $\dfrac{\partial\rho_V}{\partial T} = \alpha$ which is connected with the thermal expansion coefficient and $\dfrac{\partial \epsilon}{\partial \mu_V} = \gamma$ which could be called something like energetic susceptibility. We start with the expressions for Chiral Magnetic Effect (CME) \cite{ref:CMW}
\begin{equation}\label{CME}
\jj_V = \frac{\mu_A \BB}{2\pi^2}, 
\end{equation}
Chiral Separation Effect (CSE) \cite{ref:CSE1, ref:CSE2}
\begin{equation}\label{CSE}
\jj_A = \frac{\mu_V \BB}{2\pi^2},
\end{equation}
Chiral Vortical Effect (CVE) \cite{ref:CVE1, ref:CVE2, ref:CVE3, ref:CVE4}
\begin{eqnarray}\label{CVE}
\jj_V = \frac{\mu_V\mu_A\om}{\pi^2}, ~~~~~~\\
\jj_A = \left(\frac{T^2}{6} + \frac{\mu_V^2 + \mu_A^2}{2\pi^2}\right)\om,
\end{eqnarray}
and nondissipative energy transfer \cite{ref:CVE4,ref:En1}
\begin{equation}\label{ET}
\jj_E = \frac{\mu_v\mu_A\BB}{2\pi^2} + \frac{\mu_A\om}{3}\left(\frac{3\mu_V^2 + \mu_A^2}{\pi^2}+ T^2\right).
\end{equation}
Here $\jj_V$ is vector current, $\jj_A$ is axial current, $\jj_E$ is energy current, $\mu_V$ an $\mu_A$ are vector and axial chemical potentials, $T$ is temperature, $\BB$ is external magnetic field and $\om$ is the angular velocity of the fluid. We also use the conservation laws for vector and axial charge densities and for energy:
\begin{eqnarray} \label{Hydro:cont}
\nonumber
\partial_t \rho_V + \bm\nabla\cdot \jj_V = 0  ,\\
\partial_t \rho_A + \bm\nabla\cdot \jj_A = 0  ,\\ \nonumber
\partial_t \epsilon + \bm\nabla\cdot \jj_E = 0  .~
\end{eqnarray}
Here $\rho_V$ and $\rho_A$ are vector and axial charge densities and $\epsilon$ is energy density. Let us consider small fluctuations of these densities (and therefore the corresponding currents) above the equilibrium and denote them by $\delta\rho_V$, $\delta\rho_A$ and $\delta\epsilon$ (and the corresponding fluctuations of the currents as $\delta\jj_V$, $\delta\jj_A$, $\delta\jj_E$). They are connected with the fluctuations in chemical potentials $\delta\mu_V$, $\delta\mu_A$ and temperature $\delta T$ in the following way:
\begin{eqnarray} \label{Hydro:connect}
\nonumber
\delta \rho_V = \chi \delta\mu_V + \alpha\delta T ,\\
\delta \rho_A = \chi \delta\mu_A ,~~~\\ \nonumber
\delta \epsilon = \gamma\delta\mu_V + C_V\delta T.
\end{eqnarray}
Here $\chi$ is charge susceptibility and $C_V$ is heat capacity. This is the most general view of the connection between these fluctuations in the case of unbroken chiral symmetry, which implies $\mu_A = 0$ in the background. Indeed, $\rho_V$, $\mu_V$ , $\epsilon$ and $T$ are parity-even quantities, while $\rho_A$, $\mu_A$ are parity-odd, so terms like $\dfrac{\partial \rho_A}{\partial \mu_V}$, $\dfrac{\partial \rho_A}{\partial T}$, $\dfrac{\partial \rho_V}{\partial \mu_A}$, $\dfrac{\partial \epsilon}{\partial \mu_A}$ should be all equal to zero in our case of background $\mu_A = 0$. 

Rewriting the equations (\ref{CME}--\ref{ET}) in terms of small fluctuations above the equilibrium and keeping only the linear terms we get
\begin{eqnarray}
\nonumber
\delta\jj_V = \frac{\delta\mu_A}{2\pi^2}\BB' ,~~~~~~~~~ \\
\delta\jj_A = \frac{\delta\mu_V}{2\pi^2}\BB' +\frac{T}{3}\delta T\om , ~~~\\ \nonumber
\delta\jj_E = \frac{\mu_V\delta\mu_A}{2\pi^2}\BB' + \frac{\delta\mu_A}{3}T^2\om, 
\end{eqnarray}
where we introduced the effective magnetic field $\BB'\equiv \BB + 2\mu_V\om$. Combining these equations with the continuity equations (\ref{Hydro:cont}) and the linear connections (\ref{Hydro:connect}) and going to the Fourier components, we obtain the following system : 
\begin{equation}
\left[\nu\begin{pmatrix}
\chi & 0 & \alpha \\ 0 & \chi & 0 \\ \gamma & 0 & C_V
\end{pmatrix} +  \begin{pmatrix}
 0 & \dfrac{(\kk\cdot\BB')}{2\pi^2} & 0 \\ \dfrac{(\kk\cdot\BB')}{2\pi^2} & 0 & \dfrac{T}{3}(\kk\cdot\om) \\ 0 & \dfrac{\mu_V(\kk\cdot\BB')}{2\pi^2} + \dfrac{T^2}{3}(\kk\cdot\om) & 0
\end{pmatrix}\right] \begin{pmatrix}
\delta \mu_V  \\ \delta \mu_A \\ \delta T
\end{pmatrix} = \begin{pmatrix}
 0\\ 0 \\ 0
\end{pmatrix}.
\end{equation}
To obtain the dispersion relation we first need to multiply this equation by the inverse of the matrix accompanying $\nu$. It is natural to wonder if this matrix is non-degenerate. In our further analysis we will assume that it is, and just note now that if this matrix is degenerate (so that $\chi C_V - \alpha\gamma = 0$) then our linear approximation breaks down and we would have to take into account the terms of higher power in $\delta\mu_V,\delta\mu_A,\delta T$. So, assuming that the matrix is non-degenerate we find that there are three  branches of dispersion relation, $\nu = 0, \pm \lambda$, where $\lambda$ is the positive eigenvalue of the matrix 
\begin{equation}
\frac{1}{\chi C_V - \alpha\gamma}\begin{pmatrix}
 0& \dfrac{C_V(\kk\cdot\BB')}{2\pi^2} - \dfrac{\alpha\mu_V(\kk\cdot\BB')}{2\pi^2}  - \dfrac{\alpha T^2 (\kk\cdot\om)}{3} & 0 \\ \left(C_V - \dfrac{\alpha\gamma}{\chi}\right)\dfrac{(\kk\cdot\BB')}{2\pi^2} & 0 & \left(C_V - \dfrac{\alpha\gamma}{\chi}\right)\dfrac{T}{3}(\kk\cdot\om) \\ 0  & -\dfrac{\gamma (\kk\cdot\BB')}{2\pi^2} + \dfrac{\chi\mu_V(\kk\cdot\BB')}{2\pi^2} + \dfrac{\chi T^2 (\kk\cdot\om)}{3} & 0
\end{pmatrix},
\end{equation} 

if there is any. This eigenvalue is given by the following expression:
\begin{equation}\label{Hydro_vel_gen}
\frac{1}{\chi C_V-\alpha\gamma}\sqrt{M_{12}M_{21} + M_{23}M_{32}},
\end{equation}
where
\begin{eqnarray}
M_{12} = \frac{C_V(\kk\cdot\BB')}{2\pi^2} - \frac{\alpha\mu_V(\kk\cdot\BB')}{2\pi^2}  - \frac{\alpha T^2 (\kk\cdot\om)}{3}, \\
M_{21} = \left(C_V - \frac{\alpha\gamma}{\chi}\right)\frac{(\kk\cdot\BB')}{2\pi^2},~~~~~~~~~~~~~ \\
M_{23} = \left(C_V - \frac{\alpha\gamma}{\chi}\right)\frac{T}{3}(\kk\cdot\om),~~~~~~~~~~~~ \\
M_{32} = -\frac{\gamma (\kk\cdot\BB')}{2\pi^2} + \frac{\chi\mu_V(\kk\cdot\BB')}{2\pi^2} + \frac{\chi T^2 (\kk\cdot\om)}{3}.
\end{eqnarray}

Once again, we want to find the maximal and minimal possible values of the relation $\nu/k$ for vectors $\BB$, $\om$ and $\kk$ all lying in the same plane. Let us introduce angles : $\phi$ between $\BB$ and $\om$, $\phi1$ between $\BB$ and $\kk$ and $\phi2$ between $\om$ and $\kk$ (so $\phi_2$ = $\phi$ - $\phi_1$). Now the expression  (\ref{Hydro_vel_gen}) has the form
\begin{equation}
\frac{1}{\chi C_V - \alpha\gamma}\sqrt{a_1\cos\phi_1^2 + a_2\cos\phi_1\phi_2 + a_3\cos\phi_2^2}
\end{equation}
with the coefficients
\begin{eqnarray}
a_1 = \frac{B^2(C_V - \alpha\mu_V)(\chi C_V - \alpha\gamma)}{4\pi^4\chi} ,~~~~~~~~~~~~~~~~~~~~~~~~ \\
a_2 = \frac{B\omega(\chi C_V - \alpha\gamma)[6\mu_V(C_V-\alpha\mu_V) - \pi^2\alpha T^2 + \pi^2 T(\chi\mu_V-\gamma)]}{6\pi^4\chi} ,~~~ \\
a_3 = \frac{\omega^2(\chi C_V - \alpha\gamma)[9\mu_V^2(C_V - \alpha\mu_V) - 3\mu_V T\pi^2(\alpha T+\gamma - \chi\mu_V) + \pi^2\chi T^3]}{9\pi^4\chi}.
\end{eqnarray}
Then the solution is given by the formula analogous to (\ref{vHMV}) with the new coefficients:
\begin{equation}
v_{HMV\pm}^2 = \frac{1}{(\chi C_V - \alpha\gamma)^2}\left|\frac{a_1 + a_2\cos\phi+a_3}{2} \pm \frac{1}{2}\sqrt{a_1^2 + a_2^2 + a_3^2 +2a_2(a_1+a_3)\cos\phi+2a_1a_3\cos2\phi}~\right|.
\end{equation}
 It might be interesting to find if the maximal velocity could be equal to zero. Obviously, this is only possible if $\BB ||\om$ (note that in this case the minimal velocity is equal to zero automatically). By letting $v_{HMV+}=0$ in the above expression with $\phi=0$ we find:
 \begin{equation}
 B+2\mu_V\omega = \frac{\pi^2 T}{3(C_V-\gamma\mu_V)}\omega\left(\alpha T + \gamma -\chi\mu_V \pm \sqrt{(\alpha\ T + \gamma -\chi\mu_V)^2 - 4(C_V-\alpha\mu_V)\chi T}\right).
 \end{equation}
So we see that the solution exists only if $C_V\chi T \leq \alpha\mu_V\chi T + \frac{1}{4}(\alpha T + \gamma -\chi\mu_V)^2$ which is a condition on parameters. For example, in case of hot Weyl gas, which we have studied in kinetic theory, $\alpha\varpropto\mu_V$, $\gamma\varpropto\mu_V$(while $\chi\varpropto T^2$ and $C_V\varpropto T^3$) and $\mu_V\ll T$, so we see that the above condition is never satisfied and hence the velocity is always larger than zero. 

The expression for group velocity for arbitrary $\kk$ via $v_{HMV\pm}$ is similar to one given in \cite{ref:CHW} so we would not repeat it here and just note that directions of $\kk$ and $\dfrac{\partial \nu}{\partial\kk}$ are not the same, in general.

Finally, let us write down the expression for velocities in special cases of $\mu_V = 0$ (mixed heat and magnetic wave), $\BB = 0$, $\mu_V \neq0$ (mixed heat and vortical wave) and $\BB = 0, \mu_V =0$ (pure heat wave). We consider the most general case of $\alpha$ and $\gamma$ independent on $\mu_V$ (unlike the case of hot Weyl gas in kinetic theory), so the expressions for velocities of pure heat wave and mixed heat and magnetic wave are also different from the results obtained in \cite{ref:CHW}. For the mixed heat and magnetic wave we obtain: 
\begin{equation}
v_{MH\pm}^2 = \left|\frac{a_1 + a_2\cos\phi+a_3}{2} \pm \frac{1}{2}\sqrt{a_1^2 + a_2^2 + a_3^2 +2a_2(a_1+a_3)\cos\phi+2a_1a_3\cos2\phi}~\right|
\end{equation}
with the coefficients 
\begin{eqnarray}
a_1 = \frac{B^2C_V}{4\pi^4\chi(\chi C_V - \alpha\gamma)} ,~~ 
a_2 = -\frac{B\omega T(\alpha T +  \gamma)}{6\pi^2\chi(\chi C_V - \alpha\gamma)} ,~~
a_3 = \frac{\omega^2 T^3}{9\pi^2(\chi C_V - \alpha\gamma)}.
\end{eqnarray}
For the mixed heat and vortical wave we obtain : 
\begin{equation}
v_{HV}^2 = \left|\frac{\omega^2[9\mu_V^2(C_V - \alpha\mu_V) - 3\mu_V T\pi^2(\alpha T+\gamma - \chi\mu_V) + \pi^2\chi T^3]}{9\pi^4\chi(\chi C_V - \alpha\gamma)}~\right|.
\end{equation}
For the pure heat wave we have:
\begin{equation}
v_{CHW}^2 = \left|\frac{\omega^2T^3}{9\pi^2(\chi C_V - \alpha\gamma)}~\right|.
\end{equation}

Let us notice that these results for velocities coincide with the results obtained in section \ref{sec:Waves} in kinetic theory.

\section{Conclusions}

We have confirmed the existence of Chiral Heat Wave and its' mixings with Chiral Magnetic Wave and Chiral Vortical Wave, first suggested in \cite{ref:CHW}, in chiral kinetic theory. It was shown that the results for velocities of Chiral Heat Wave and mixed Chiral Magnetic-Heat Wave obtained in kinetic theory coincide with the ones obtained in \cite{ref:CHW} while the velocities of mixed Chiral Heat-Vortical Wave and mixed Chiral Heat-Magnetic-Vortical Wave are slightly different, as in \cite{ref:CHW} were not taken into account the non-zero, in general, values of $\dfrac{\partial\rho_V}{\partial T}$ and $\dfrac{\partial\epsilon}{\partial \mu_V}$. Because of that, we also presented the corrected expressions for these velocities in hydrodynamic approach.

There remain plenty of uncovered topics related to this subject, for example how such waves would look like in the opposite, low-temperature, high-density limit in kinetic theory. It might be also interesting to investigate the transition between these waves and Landau's zero sound (similar analysis was made in \cite{ref:Kinetic1} for Chiral Magnetic Wave). Hopefully, these questions would be answered in the future studies.

\section{Acknowledgments}

Author is indebted to A. S. Gorsky for suggesting this problem and numerous valuable discussions. This work was supported in part by grant RFBR-15-02-02092.

\begin{thebibliography}{99} 

\bibitem{ref:CME} 
  K.~Fukushima, D.~E.~Kharzeev and H.~J.~Warringa,
{\sl ``The Chiral Magnetic Effect''},
  Phys.\ Rev.\ D {\bf 78}, 074033 (2008) [arXiv:0808.3382].
\bibitem{ref:CSE1}
  D.~T.~Son and A.~R.~Zhitnitsky,
{\sl ``Quantum anomalies in dense matter''},
  Phys.\ Rev.\ D {\bf 70}, 074018 (2004)
  [hep-ph/0405216].
\bibitem{ref:CSE2}
  M.~A.~Metlitski and A.~R.~Zhitnitsky,
  {\sl ``Anomalous axion interactions and topological currents in dense matter''},
  Phys.\ Rev.\ D {\bf 72}, 045011 (2005) [hep-ph/0505072].
\bibitem{ref:CVE1} 
N.~Banerjee, J.~Bhattacharya, S.~Bhattacharyya, S.~Dutta, R.~Loganayagam and P.~Surowka,
{\sl ``Hydrodynamics from charged black branes''},
JHEP {\bf 1101}, 094 (2011)
[arXiv:0809.2596].
\bibitem{ref:CVE2} 
  J.~Erdmenger, M.~Haack, M.~Kaminski and A.~Yarom,
  {\sl ``Fluid dynamics of R-charged black holes''},
  JHEP {\bf 0901}, 055 (2009)
  [arXiv:0809.2488].
\bibitem{ref:CVE3} 
  D.~T.~Son and P.~Surowka,
  {\sl ``Hydrodynamics with Triangle Anomalies''},
  Phys.\ Rev.\ Lett.\  {\bf 103}, 191601 (2009) 
  [arXiv:0906.5044].
\bibitem{ref:CVE4} 
K.~Landsteiner, E.~Megias and F.~Pena-Benitez,
{\sl ``Gravitational Anomaly and Transport''},
Phys.\ Rev.\ Lett.\  {\bf 107}, 021601 (2011)  
[arXiv:1103.5006].
\bibitem{ref:CMW}
  D.~E.~Kharzeev and H.~U.~Yee,
  {\sl ``Chiral Magnetic Wave''},
  Phys.\ Rev.\ D {\bf 83}, 085007 (2011)
  [arXiv:1012.6026].
\bibitem{ref:CVW}
  Y.~Jiang, X.~G.~Huang and J.~Liao,
{\sl ``Chiral vortical wave and induced flavor charge transport in a rotating quark-gluon plasma''},
  arXiv:1504.03201 [hep-ph].
\bibitem{ref:CHW}
  M.~N.~Chernodub,
  {\sl ``Chiral Heat Wave and wave mixing in chiral media''},
  arXiv:1509.01245 [hep-th].
\bibitem{ref:Kinetic1} 
  M.~A.~Stephanov, H.~U.~Yee and Y.~Yin,
  {\sl ``Collective modes of chiral kinetic theory in a magnetic field''},
  Phys.\ Rev.\ D {\bf 91}, 125014 (2015) [arXiv:1501.00222].
\bibitem{ref:Kinetic2}  
  M.~A.~Stephanov, Y.~Yin,
  {\sl ``Chiral Kinetic Theory''}
  Phys. Rev. Lett. {\bf 109},162001 (2012)[arXiv:1207.0747]
\bibitem{ref:En1}
  K.~Landsteiner, E.~Megias, L.~Melgar and F.~Pena-Benitez,
{\sl ``Holographic Gravitational Anomaly and Chiral Vortical Effect''},
  JHEP {\bf 1109}, 121 (2011) [arXiv:1107.0368].
\bibitem{ref:Kin_my}
  D.~ Frenklakh, 
  {\sl ``Chiral Magnetic-Vortical Wave''}
  arXiv:1510.09198 [hep-th]
\end
{thebibliography}

\end{document}